\documentclass[12pt]{article}
\usepackage{a4wide}
\pdfoutput=1

\usepackage[utf8]{inputenc}
\usepackage{draft}
\usepackage{putex}
\usepackage[all]{xy}
\usepackage{stackrel,amssymb}
\usepackage{mathtools}
\usepackage{graphicx}
\usepackage{hhline}
\usepackage{cellspace}
\usepackage{caption}
\usepackage{dsfont}
\usepackage{amsfonts}
\usepackage{amsmath}
\usepackage{amsthm}
\usepackage{bm}
\usepackage{array}
\usepackage{subcaption}
\usepackage{epstopdf}
\usepackage{enumerate}
\usepackage{cite}
\usepackage{tensor}
\usepackage{slashed}
\usepackage{rotating}
\usepackage{multirow}
\usepackage{longtable}
\usepackage{ytableau}
\usepackage{tikz,tikz-3dplot}
\usepackage{tikz-cd}
\usetikzlibrary{shapes.misc,shadows,fit,decorations.pathmorphing,positioning,trees,decorations.markings,decorations.pathreplacing,calc,shapes,patterns,arrows}
\usepackage{environ}
\usepackage{listings}
\usepackage{subfiles}
\usepackage{tabularx}
\usepackage{footnote}
\usepackage{etoolbox}
\usepackage{thmtools}
\usepackage[framemethod=Tikz]{mdframed}
\usepackage{comment}
\usepackage[normalem]{ulem}
\usepackage{placeins}
\usepackage{makecell}
\usepackage[most]{tcolorbox}

\usepackage{xparse}
\usepackage{hyperref}
\usepackage[capitalize]{cleveref}
\usepackage{pdflscape}
\usepackage{everypage}

\definecolor{greenZS}{RGB}{10,153,33}
\definecolor{redZS}{RGB}{255,92,0}
\definecolor{blueZS}{RGB}{44,96,241}
\definecolor{grayZS}{RGB}{235,235,235}
\definecolor{orangeZS}{RGB}{255,133,0}

\crefname{definition}{Def.}{Defs.}

\allowdisplaybreaks[1] 

\ytableausetup{smalltableaux,centertableaux}

\hypersetup{
pdftitle={On generalized symmetries for disconnected gauge groups},    
pdfauthor={\textcopyright\ Guillermo Arias Tamargo, Diego Rodríguez Gómez},     
pdfsubject={HEP},   
pdfcreator={pdfLaTex},   
pdfproducer={LaTex}, 
pdfkeywords={},
colorlinks=true,
linkcolor=blue,
urlcolor=blue,
filecolor=blue,
citecolor=blue,
linktocpage=true
}

\numberwithin{equation}{section}
\newcommand{\sz}{\text{sz}}
\newcommand{\ZZ}{\mathbb{Z}}

\begin{document}

\begin{flushright}\footnotesize

\texttt{}
\vspace{0.6cm}
\end{flushright}

\mbox{}
\vspace{0truecm}
\linespread{1.1}

\centerline{\LARGE \bf Non-Invertible Symmetries from Discrete Gauging}
\medskip
\medskip
\centerline{\LARGE \bf  and Completeness of the Spectrum }
\medskip

\vspace{1.5truecm}

\centerline{
    { Guillermo Arias-Tamargo}\footnote{guillermo.arias.tam@gmail.com}, and
    { Diego Rodr\'iguez-G\'omez}\footnote{d.rodriguez.gomez@uniovi.es}}

\vspace{1cm}
\centerline{{\it  Department of Physics, Universidad de Oviedo}} \centerline{{\it C/ Federico Garc\'ia Lorca  18, 33007  Oviedo, Spain}}
\medskip
\centerline{{\it  Instituto Universitario de Ciencias y Tecnolog\'ias Espaciales de Asturias (ICTEA)}}\centerline{{\it C/~de la Independencia 13, 33004 Oviedo, Spain.}}
\medskip
\vspace{1cm}

\centerline{\small{\bf Abstract} }

\begin{center}
\begin{minipage}[h]{\textwidth}
We study global 1- and $(d-2)$-form symmetries for gauge theories based on disconnected gauge groups which include charge conjugation. For pure gauge theories, the 1-form symmetries are shown to be non-invertible. In addition, being the gauge groups disconnected, the theories automatically have a $\mathbb{Z}_2$ global $(d-2)$-form symmetry. We propose String Theory embeddings for gauge theories based on these groups. Remarkably, they all automatically come with twist vortices which break the $(d-2)$-form global symmetry. This is consistent with the conjectured absence of global symmetries in Quantum Gravity.
\end{minipage}

\end{center}
\newpage

\tableofcontents 

\section{Introduction}

Global symmetries play a central role in Quantum Field Theory (QFT). They are used as an organizing principle to systematically construct the possible operators, their breaking pattern allows to characterize the phases of a system and their possible anomalies provide exact constraints on the dynamics. However, in recent times it has been made clear that the notion of symmetry has to be generalized from the traditional textbook definition typically in terms of Noether currents. The central idea, pioneered in \cite{Gaiotto:2014kfa}, is that symmetries are associated to symmetry operators $T_g(M^{d-(p+1)})$ depending on a transformation $g$ and defined on codimension $p+1$ manifolds $M^{d-(p+1)}$.\footnote{We will mostly assume the QFT defined on a dimension $d$ euclidean signature space, and thus talk about operator insertions.} The crucial point is that the dependence on $M^{d-(p+1)}$ is topological: the properties of $T_g(M^{d-(p+1)})$ --for instance their correlation functions-- do not change under small changes of $M^{d-(p+1)}$ as long as these do not cross any charged operator. 

The textbook examples of global symmetries naturally fit in this framework. Indeed, for a continuous symmetry there is a Noether current, whose integral on $M^{d-1}$ manifolds gives a charge $Q$. Clearly, slight changes of $M^{d-1}$ do not change $Q$ as long as these do not cross charged operators. Moreover, the exponential of $Q$ gives a an element of the symmetry group, and thus corresponds to the $T_g(M^{d-1})$.\footnote{In this case, as $e^{iQ}$ is a group element, it is often denoted by $U_g(M^{d-1})$.} The point of view above naturally generalizes this in two directions. On one hand it allows for more generic symmetries supported on codimension $p+1$ manifolds whose charged objects are supported on $p$-dimensional submanifolds. These are often referred to as \textit{higher form symmetries} or \textit{p-form symmetries}. On the other hand, it allows to consider more generic categorical symmetries not arising from a group. This is reflected into a more generic fusion rule for symmetry operators, which in particular do not need to have an inverse (as opposed to what should happen for a fusion rule of elements in a group). These cases are often dubbed \textit{non-invertible symmetries}.

The existence of non-invertible symmetries is well-known in lower-dimensional QFT's. In particular, in 2d there is a whole body of work studying these (see \textit{e.g.} \cite{Verlinde:1988sn,Petkova:2000ip,Fuchs:2002cm,Bachas:2004sy,Fuchs:2007tx,Bachas:2009mc} for early references). Their status in higher dimensions is however a bit less clear.\footnote{Indeed, \cite{Casini:2020rgj,Casini:2021zgr} claim that these do not exist in $d>3$.} The case of $O(2)$ has been argued to give rise to non-invertible symmetries in \cite{Heidenreich:2021xpr}, and more exotic examples have been constructed in \cite{Choi:2021kmx,Kaidi:2021xfk}. More recently, it has been argued in \cite{Roumpedakis:2022aik} that indeed non-invertible symmetries are common in higher dimensions.

Very recently, \cite{Heidenreich:2021xpr} (see also \cite{Rudelius:2020orz}) provided a criterion to compute the symmetry operators in a gauge theory, including both invertible (\textit{i.e.} usual symmetries associated to groups) as well as non-invertible symmetries. From the analysis in \cite{Heidenreich:2021xpr}, it follows that if a gauge theory (whose gauge group we assume to be compact) has local operators in all possible representations, no possible non-trivial topological operator candidate for electric 1-form symmetry operator can exist.\footnote{By electric 1-form symmetry we mean the one that is always present in a gauge theory, associated to the field strenght. In the free Maxwell case, the Noether current is simply $F$.} Therefore, the absence of global electric 1-form symmetries is equivalent to the completeness of the spectrum of the QFT.\footnote{Completeness of the spectrum is defined as the existence of operators in every possible representation of the gauge group.} In turn, this has interesting implications for the Swampland Program (in short, the study of the restrictions imposed in the low energy Physics which can be consistently coupled to Quantum Gravity. See \textit{e.g.} \cite{Palti:2019pca,vanBeest:2021lhn} for introductions and further references), where the Absence of Global Symmetries and the Completeness of the Spectrum are two central conjectures which indeed have been long suspected to be deeply related. 

In this paper we study in detail (certain) higher-form global symmetries of gauge theories which include, as an element of the gauge group, charge conjugation in generic $d$ dimensions (we stress that there may be particularities for given $d$'s which we leave for future study). More precisely, we will consider gauge theories based on the gauge groups constructed in \cite{Bourget:2018ond,Arias-Tamargo:2019jyh} dubbed $\widetilde{SU}(N)$. These are principal extensions of $SU(N)$ by the $\mathbb{Z}_2$ outer automorphism corresponding to flipping the Dynkin diagram, which, in particular, exchanges the fundamental representation with the antifundamental, and thus corresponds to charge conjugation (the construction can be extended to $U(N)$, giving rise to $\widetilde{U}(N)$). Concentrating on pure gauge theories, we will study the 1-form electric symmetry, which turns out to be non-invertible (in a sense, generalizing the $O(2)$ example). Moreover, as the gauge groups are disconnected, there is a $(d-2)$-form symmetry associated to the non-trivial $\pi_0(G)$ for $G=\widetilde{SU}(N),\,\widetilde{U}(N)$.\footnote{There may be other higher form symmetries other than those we consider. In particular, there may be a $(d-3)$-form magnetic symmetry associated to the dual of the gauge field. However, its study requires the knowledge of the GNO dual group, which is not known at present for the theories at hand.} We also introduce String Theory constructions of these theories. Amusingly, these automatically all come with configurations of extended objects which break the $(d-2)$-form symmetry. From this perspective, they may be regarded as Swampland examples in the sense that when the gauge theory with gauge group $G$ is embedded into a consistent theory of Quantum Gravity, the otherwise present $(d-2)$-form symmetry is broken by the presence of charged ``matter" (in this case extended objects).

Te remainder of this paper is structured as follows. In section \ref{sec:review} we review basic facts in the topic of higher form global symmetries, mainly from \cite{Gaiotto:2014kfa} and recent progress in \cite{Heidenreich:2021xpr}. In section \ref{section:electric 1-form} give a lightning review of the groups $\widetilde{SU}(N),\,\widetilde{U}(N)$ and study the electric 1-form symmetries of pure gauge theories based on them. In section \ref{section:d-2 form} we study the $(d-2)$-form symmetry coming from the fact that the groups are disconnected. We discuss the would-be charged objects, which are the so-called \textit{Alice strings} \cite{Schwarz:1982ec} (or \textit{twist vortices} in the nomenclature of \cite{Heidenreich:2021xpr}). As it is well-known, in the presence of twist vortices, only a subgroup of the gauge group is globally well-defined \cite{Alford:1989ch}. We also introduce a stringy construction for gauge theories based on $\widetilde{U}(N)$, which, as advertised above, automatically come with Alice strings which break the $(d-2)$-global symmetry.

\vspace{.5cm}
\textbf{Note added}: as this note was being finished, \cite{Bhardwaj:2022yxj} appeared overlapping with our results on the electric 1-form (generically non-invertible) symmetries of the $\widetilde{SU}(N)$ theories.

\section{Higher form symmetries and topological operators}\label{sec:review}

In the quest to generalize the notion of symmetry to higher-form global symmetries \cite{Gaiotto:2014kfa}, one quickly realizes that the usual textbook formulation, based on a lagrangian and a explicit transformation of the fields, is not appropriate. Instead, the focus should be on the symmetry generators $U_g(M^{d-1})$ depending on a symmetry transformation $g$ and associated to a manifold $M^{d-1}$. In the continous case, these are given by the exponentiation of the charge computed as the integral of the Noether current,
\begin{align}
    Q(M^{d-1})=\int_{M^{d-1}}\star J\,,
\end{align}
The key is that the dependence of $U_g$ on the manifold $M^{d-1}$ in which it is supported is topological: $U_g$ doesn't change under deformations of $M^{d-1}$ unless the deformation crosses an operator charged under the symmetry.

This point of view can be easily generalized to higher-form symmetries. The symmetry operators now live on a codimension $p+1$ manifold (on whom they depend only topologically), and the charged objects are extended on $p$ spatial dimensions. 

Usually, the symmetry transformations form a group,
\begin{align}\label{eq:invertible_hfs}
    U_{g_1}(M^{d-p-1}) \cdot U_{g_2}(M^{d-p-1})=U_{g_1 g_2}(M^{d-p-1})\,,
\end{align}
and the transformation has an inverse $U_g^{-1}(M^{d-p-1}) =U_{g^{-1}}(M^{d-p-1})$. However, this requirement can be relaxed, by demanding instead that the topological operators fuse according to (we now denote the operators by $T$ to stress that they may not come from a group)
\begin{align}
    T_{a}(M^{d-p-1}) \cdot T_{b}(M^{d-p-1})=\sum_i N_{ab}^i\, T_{i}(M^{d-p-1})\,,
\end{align}
and need not have an inverse; this structure is that of a fusion algebra. In this case we have what is called a \emph{categorical symmetry} or \emph{non-invertible symmetry}. 

The action of the topological operators on the charged objects $O(\mathcal{C}^p)$ can be understood by introducing the symmetry operator on a sphere $S^{d-p-1}$ that surrounds $\mathcal{C}^{p}$, and then shrinking that sphere to a point, finding
\begin{align}
    T_a(S^{d-p-1}) O(\mathcal{C}^{p}) = B_O(a) O(\mathcal{C}^p)\,,
\end{align}
where $B_O(a)$ is called the linking coefficient. As an example, we can consider the electric 1-form symmetry of a gauge theory. The charged operators are the Wilson lines $W_\rho(\gamma^1)$, with $\rho$ a representation of the gauge group; and the symmetry operators, which we denote $T_a(M^{d-2})$, are the so called Gukov-Witten operators \cite{Gukov:2006jk,Gukov:2008sn}, which are labelled by a conjugacy class $a$ of the gauge group. The linking coefficient in this case is obtained from the Aharonov-Bohm interaction between the line and the codimension 2 operator \cite{Alford:1992yx,Heidenreich:2021xpr},
\begin{align}\label{eq:linking_GW_wilson_lines}
   B_{W_\rho}(a) = \frac{\chi_\rho(a)}{\text{dim } \rho}\sz (a)\,,
\end{align}
where $\chi_\rho(a)$ is the character of the representation $\rho$ evaluated in the conjugacy class of $a$, $\sz (a)$ is the number of elements of the group inside said conjugacy class and dim $\rho$ is the dimension of the representation.

In \cite{Heidenreich:2021xpr}, the question was addressed of whether or not a Gukov-Witten operator can be topological (i.e. if it generates a, possibly non-invertible, 1-form global symmetry) if it links with an endable Wilson line. The argument is as follows: consider a gauge theory with matter fields in a representation $R$. Then the Wilson lines corresponding to the representation $R$ and tensor products thereof can end and break into segments. Suppose that a GW operator 1) is topological and 2) links non-trivially with the Wilson line (i.e. the linking coefficient is different from its linking with the identity operator, $B_W(a)\neq B_1(a)$). Then, given the topological nature of the GW operator, we can consider either shrinking it on top of the Wilson line, which produces the linking coefficient $B_W(a)$; or breaking the Wilson line into segments and shrinking the GW on top of a point where there is no Wilson line, which produces a trivial linking $B_1(a)$. By comparison, it follows that if a GW operator links non-trivially with an endable Wilson line, it cannot be topological. Equivalently, a necessary condition for GW operators to be topological is to link trivially with endable Wilson lines.

In fact, in \cite{Heidenreich:2021xpr} it was also argued that for gauge theories this necessary condition is also sufficient: one can move on the Coulomb branch\footnote{One can imagine adding adjoint fields to Higgs the gauge group, as these will not render extra Wilson lines endable.} where the gauge group is $U(1)^r$ (and possibly some discrete factor). In that case, it is known that all GW operators that link trivially with a Wilson line are topological and can be seen to precisely coincide with those selected by the necessary condition above. Since when going back to the origin of the Coulomb branch one expects the survival of these operators, which already exhaust all the \textit{a priori} possible ones, one concludes that the criterion above is actually sufficient.

Let us consider pure gauge theories with a gauge group $G$ that is disconnected. The endable Wilson lines will correspond to the adjoint representation and its tensor products. In this case, the previous argument, together with \eqref{eq:linking_GW_wilson_lines}, leads to a very simple criterium to find the 1-form symmetry. Instead of the center of the group (as is the case in the more usual examples of connected and simply connected groups like $SU(N)$), the topological Gukov-Witten operators will correspond to the conjugacy classes of the elements in the centralizer of the identity component $G^0$ of $G$,
\begin{align}\label{eq:1-form_symmetry_general}
    \{\text{topological GW}\}\equiv \{G^{-1} h G\,,\, h\in C_{G}\left(G^{0}\right)\}\,,
\end{align}

Once the topological operators have been identified, we need to determine whether they generate a group or a non-invertible symmetry. A possible way to do it is by using the so called \emph{quantum dimension} of the operator, which is defined \cite{Heidenreich:2021xpr} as the linking coefficient with the identity operator, $\text{dim}(U_a)=B_1(a)$. As an example, if we are concerned with the one-form symmetry, the topological operators are Gukov-Witten operators and their quantum dimension is \eqref{eq:linking_GW_wilson_lines}
\begin{align}
    \text{dim}(T_a)=B_1(a)=\sz(a)\,.
\end{align}
The quantum dimensions have the property that they get multiplied under the fusion of topological operators, and summed under their sum,
\begin{align} \label{eq:quantum_dimension}
    &\dim(T_a\cdot T_b)=\dim(T_a) \dim(T_b)\,,\\
    &\dim(T_a + T_b)=\dim(T_a) + \dim(T_b)\,,
\end{align}
and since the topological operator corresponding to the identity always has quantum dimension equal to 1, this allows us to infer when a symmetry has to be non-invertible from the presence of topological operators with quantum dimension $\ge 1$.

In the same way that we can study the electric one-form symmetry from the topological GW operators and under which Wilson lines are charged, we can also look at the topological Wilson lines to find out about the dual $(d-2)$-form symmetry under which GW operators are charged. It turns out that this problem has a more straight-forward solution \cite{Heidenreich:2021xpr}. Since the gauge holonomy along a contractible loop always belongs to the identity component of the gauge group, two Wilson lines along homotopic paths differ at most by an element of $G^0$. Therefore, only Wilson lines corresponding to representations that map $G^0$ to the identity are topological. In other words,
\begin{align}\label{eq:dual_symmetry_general}
    \{\text{topological WL}\} \equiv \{ \text{representations of } \pi_0(G)\}\,,
\end{align}
where $\pi_0(G)$ is the group of connected components of $G$. Note that this discussion is actually unchanged in the presence of matter fields in any representation of $G$.

While the main focus of this work lies in pure gauge theories, one can consider more general theories adding matter fields.\footnote{Depending on the matter content, there may be gauge anomalies, as recently studied in \cite{Henning:2021ctv}.} If the matter is in a representation smaller than the adjoint, the corresponding Wilson lines will become endable, and the GW operators with whom they link can no longer be topological. As a consequence, the 1-form symmetry will be reduced. Similarly, for the dual $(d-2)$-form symmetry, we can make the GW operators endable, albeit in this case by adding suitable codimension-3 objets. These were called \emph{twist vortices} in \cite{Heidenreich:2021xpr} and are defined by a monodromy in $G/G^0$ when going around them. For the disconnected gauge groups under study in this article, these are also known as \emph{twist vortices}, or \emph{Alice strings} in 4 dimensions \cite{Schwarz:1982ec}.  

\subsection{Examples: SO, Sp, O}

In this section, we review the higher form symmetries of the orthogonal group $O(N)$ that result from using the formalism discussed above. We also list the results for $SO(N)$ and $Sp(N)$, as we will need them on latter sections.\\

\textbf{Special orthogonal and symplectic groups:} These groups are connected and simply connected, therefore the 1-form symmetry of the corresponding pure gauge theory is simply given by the center (see Table \ref{tab:GW_operators_SpSO_summary}). In all these cases, the 1-form symmetry is invertible.

\begin{table}[h!]
    \centering
    \begin{tabular}{c|c|c|c}
        Gauge group & Topological GW operators & Quantum dimension & 1-form symmetry \\\hline
        \multirow{2}{*}{$Sp(N)$} & $T^{Sp}_0=$ Id & 1  & \multirow{2}{*}{$\ZZ_2$} \\
                              & $T^{Sp}_\pi$   &  1 & \\\hline
        \multirow{2}{*}{$SO(2)$ }  &   $T^{SO(2)}_0=$ Id   &  1  & \multirow{2}{*}{$SO(2)$} \\
           &   $T^{SO(2)}_\theta\,,$ $\theta\in (0,2\pi)$ &  1  & \\\hline
        \multirow{2}{*}{$SO(2k),\,k\ge 2$ }  &   $T^{SO}_0=$ Id   &  1  & \multirow{2}{*}{$\ZZ_2$} \\
           &   $T^{SO}_\pi$ &  1  & \\\hline
         $SO(2k-1)\,,$ $k\ge2$  & $T^{SO}_0=$ Id  & 1 &  Trivial
           
    \end{tabular}
    \caption{Summary of topological Gukov-Witten operators for theories with $Sp(N)$ and $SO(N)$ gauge group.}
    \label{tab:GW_operators_SpSO_summary}
\end{table}

The magnetic $(d-2)$-form symmetry, which is given by the group of connected components, is trivial in all these cases. 

\textbf{Orthogonal groups:} This is the first instance of disconnected gauge group that we encounter. Instead of the center, the 1-form symmetry is obtained from the centralizer of the identity component of the group. We need to distinguish three possible cases: $N=2$, $N$ even and bigger than 2, or $N$ odd.

The case of $O(2)$ was studied in detail in \cite{Heidenreich:2021xpr}, and it is special because the identity component, $SO(2)$, is abelian. The full group $O(2)$ can be written as a semidirect product $ SO(2)\rtimes\ZZ_2$. By definition the generator of the $\ZZ_2$ does not commute with the $SO(2)$, therefore the centralizer is
\begin{align}
    C_{O(2)}(SO(2))=SO(2)\,.
\end{align}

The topological GW operators are labelled by the conjugacy classes of elements in this centralizer. Here the global structure of the group becomes relevant, as we can also conjugate by the nontrivial element in the $\ZZ_2$. If we denote this element as $P$, the action on an element of the centralizer is
\begin{align}
   P^{-1}\cdot \left( \begin{array}{cc}
        \cos \theta & - \sin \theta \\
        \sin \theta & \cos \theta
    \end{array}\right)  \cdot P = \left( \begin{array}{cc}
        \cos \theta &  \sin \theta \\
        -\sin \theta & \cos \theta
    \end{array}\right) \,,
\end{align}
i.e. it maps $\theta\mapsto -\theta$. This means that we don't have one GW operator for each $\theta\in [0,2\pi]$, but rather one for each $\theta\in[0,\pi]$ and the quantum dimension of the operators labelled by $\theta\in(0,\pi)$ is equal to two. Therefore, the 1-form symmetry in the $O(2)$ case is non-invertible. The fusion algebra of the topological operators was reported in \cite{Heidenreich:2021xpr},
\begin{align}\label{eq:fusion_algebra_O2}
   & T^{O(2)}_{\theta}\cdot T^{O(2)}_{\varphi}  = T^{O(2)}_{\theta+\varphi} + T^{O(2)}_{\theta-\varphi}\,,\nonumber\\
&    T^{O(2)}_{\theta} \cdot T^{O(2)}_\pi   = T^{O(2)}_{\theta+\pi}\,,\nonumber\\
 &   T^{O(2)}_{\pi} \cdot T^{O(2)}_{\pi}  = 1 \,,\\
  &  T^{O(2)}_\theta \cdot T^{O(2)}_{\theta} = 1 + W^{O(2)}_{\text{sign}} + T^{O(2)}_{2\theta}\,, \nonumber\\
   & T^{O(2)}_{\theta} \cdot T^{O(2)}_{\pi-\theta} = T^{O(2)}_\pi + W^{O(2)}_{\text{sign}} T^{O(2)}_{\pi} + T^{O(2)}_{2\theta-\pi}\,,\nonumber
\end{align}
where $\theta\neq\varphi$ and $W^{O(2)}_{\text{sign}}$ is the Wilson line in the sign representation of $O(2)$. The appearance of the Wilson line in the fusion of two GW operators is the hallmark of a higher-group global symmetry structure, that can also be seen from the fact that Wilson lines are charged under the (zero-form) charge conjugation symmetry of $SO(2)$. In more detail, the fourth equation in \eqref{eq:fusion_algebra_O2} can be understood as follows.\footnote{We thank Miguel Montero for explaining this argument to us.} First, consider the fusion of two GW operators corresponding to different angles $\theta$ and $\varphi$ and take the limit $\varphi\to\theta$. We obtain
\begin{align}\label{eq:O2_weird_fusion_1}
    T_\theta^{O(2)}\cdot T_\theta^{O(2)} = T_{2\theta}^{O(2)} + \text{`` }T_0^{O(2)}\text{ ''}\,.
\end{align}
Naively, one would say that $\text{`` }T_0^{O(2)}\text{ ''}$ is equal to two copies of the identity. However, to properly investigate this one should consider the fusion inside correlation functions. When all other operators in the correlator belong to the connected component (that is, they are operators just like those in the $SO(2)$ theory), indeed $T_0^{O(2)}$ looks like twice the identity. However, if one of the inserted operators belongs to the disconnected component there may be subtleties. Indeed, suppose including in our correlator the GW operator corresponding to the $\ZZ_2\subseteq O(2)$, which we denote $T^{O(2)}_{\text{disc}}$. This corresponds to the insertion of an Alice string defined by a gauge connection which picks a sign upon going around the string. Suppose now $\langle T_{\rm disc}^{O(2)}T_{\theta}^{O(2)}\cdots\rangle$. Since $T_{\theta}^{O(2)}$ shifts the gauge connection by a constant, which is clearly incompatible with the action of $T_{\rm disc}^{O(2)}$, it follows that $T_{\rm disc}^{O(2)}T_{\theta}^{O(2)}=0$ for any $\theta$. Thus, inserting \eqref{eq:O2_weird_fusion_1} in the correlator with $T_{\rm disc}^{O(2)}$ leads to the requirement $\text{`` }T_0^{O(2)}\text{ ''}T_{\rm disc}^{O(2)}=0$, which shows that ``$T_0^{O(2)}$'' cannot simply be two copies of the identity. In fact, the only operator we can construct that satisfies these conditions is
\begin{align}\label{eq:O2_weird_fusion_2}
    \text{`` }T_0^{O(2)}\text{ ''} = 1+ W^{O(2)}_{\rm sign}\,,
\end{align}
leading to the fusion rule in \eqref{eq:fusion_algebra_O2} (a similar argument would hold for the fifth equation).

The cases of $O(N)$ where $N\ge 3$ are simpler, since the centralizer is always finite. If $N$ is odd, then $SO(N)$ has trivial center. However, in this case $O(N)$ is the direct product $ SO(N)\times\ZZ_2$ and the nontrivial element of the $\ZZ_2$ (which is $-\mathds{1}$) will appear in the centralizer. Naturally, both $\pm\mathds{1}$ are mapped to themselves by conjugation in $O(N)$, thus, we have a $\ZZ_2$ invertible 1-form symmetry. On the other hand, if $N$ is even, $SO(N)$ has a center isomorphic to $\ZZ_2$, where the non-trivial element is precisely $-\mathds{1}$. In this case the extension to the orthogonal group is a semidirect product $ SO(N)\rtimes \ZZ_2$, and no new elements will appear in the centralizer. We conclude that also in this case we have an invertible $\ZZ_2$ 1-form symmetry. We summarize these results in Table \ref{tab:GW_operators_O_summary}.

\begin{table}[h!]
    \centering
    \begin{tabular}{c|c|c|c}
        Gauge group & Topological GW operators & Quantum dimension & 1-form symmetry \\\hline
        \multirow{3}{*}{$O(2)$} &  Id & 1  & \multirow{3}{*}{\eqref{eq:fusion_algebra_O2}} \\
                              & $T^{O(2)}_\theta\,,$ $\theta\in (0,\pi)$ & 2 &  \\
                              & $T^{O(2)}_\pi$   &  1 & \\\hline
        \multirow{2}{*}{$O(N),\,N\ge 3$ }  &  Id   &  1  & \multirow{2}{*}{$\ZZ_2$} \\
           &   $T^{O(N)}_\pi$  &  1  &
    \end{tabular}
    \caption{Summary of topological Gukov-Witten operators for theories with $O(N)$ gauge group.}
    \label{tab:GW_operators_O_summary}
\end{table}

The dual $(d-2)$-form symmetry is obtained from the representations of the group of connected components \eqref{eq:dual_symmetry_general}. For all the $O(N)$ groups, $\pi_0(O(N))=\ZZ_2$, which has two representations. These two representations of $\ZZ_2$ lift to the full $O(N)$ group giving rise to the trivial and sign representation. Therefore, the topological Wilson lines are precisely $W^{O(N)}_\mathbf{1}(\gamma)$ and $W^{O(N)}_{\text{sign}}(\gamma)$. Note that in this case the fusion of the Wilson lines (which in general is obtained from the decomposition of the tensor product of the two initial representations) reduces to a group operation, 
\begin{align}\label{eq:fusion_WL_O_groups}
    W^{O(N)}_\mathbf{1}\cdot W^{O(N)}_\mathbf{1} = W^{O(N)}_\mathbf{1}\,,\qquad W^{O(N)}_\mathbf{1}\cdot W^{O(N)}_{\text{sign}} = W^{O(N)}_{\text{sign}}\,,\nonumber\\
    W^{O(N)}_{\text{sign}}\cdot W^{O(N)}_\mathbf{1} = W^{O(N)}_{\text{sign}}\,,\qquad W^{O(N)}_{\text{sign}}\cdot W^{O(N)}_{\text{sign}} = W^{O(N)}_\mathbf{1}\,,
\end{align}
and so the $(d-2)$-form symmetry is an invertible $\ZZ_2$.

\section{Electric 1-form symmetry}\label{section:electric 1-form}

In this section, we look at the electric 1-form symmetry of disconnected groups built as $\ZZ_2$ extensions of $SU(N)$ or $U(N)$. We derive the topological Gukov-Witten operators from the generic arguments presented in section \ref{sec:review}, namely from the computation of the centralizer of the identity component of these groups. An alternative derivation of which are the topological GW operators, with equal results, is presented in appendix \ref{sec:appendix_Wendt}.

We begin by recalling some basic definitions and propierties of $\widetilde{SU}(N)$ groups. These are the principal extensions of $SU(N)$ groups, i.e. semidirect products $SU(N) \rtimes_{\Theta} \mathbb{Z}_2$ where $\Theta : \mathbb{Z}_2 \rightarrow \mathrm{Aut}(SU(N))$ is a lift to the group of the automorphism of the Dynking diagram of $SU(N)$. If $N$ is odd, there is only one such possible lift up to isomorphism,
\begin{align}\label{eq:semidirect_theta1}
    \Theta^I(1) (g) = g\,,\quad \Theta^I(-1) (g) = (g^{-1})^T = \overline{g}\,,
\end{align}
where $g \in SU(N)$ and the bar denotes complex conjugation. If $N$ is even, there are two distinct choices of $\Theta$ that give rise to two different groups \cite{Arias-Tamargo:2019jyh}. One is given by \eqref{eq:semidirect_theta1} and the other by
\begin{align}\label{eq:semidirect_theta2}
    \Theta^{II}(1) (g) = g\,,\quad \Theta^{II}(-1) (g) = -J_N (g^{-1})^T J_N  =  -J_N  \overline{g}  J_N\,,
\end{align}
where 
\begin{align}
    J_{2k} := \left(\begin{matrix} 0 & -\mathds{1}_{k \times k} \\
    \mathds{1}_{k \times k} & 0 \\
    \end{matrix} \right) \, \ .
\end{align}

We denote these two different groups as $\widetilde{SU}(N)_I$ and $\widetilde{SU}(N)_{II}$ respectively, and their elements are pairs $(g,\eta)$ with $g\in SU(N)$ and $\eta\in\ZZ_2$. According to the definition of semidirect product, multiplication of elements is given by
\begin{align}
    (g_1,\eta_1) (g_2,\eta_2) = (g_1 \Theta(\eta_1)(g_2), \eta_1 \eta_2)\,.
\end{align}
It is possible to give an matrix construction of the groups explicitly exhibiting these properties \cite{Arias-Tamargo:2019jyh}.

Note that we can apply the same construction beginning with $g_i\in U(N)$, although in this case we cannot call them principal extensions. We will denote these groups as  $\widetilde{U}(N)_I$ and $\widetilde{U}(N)_{II}$. In many cases, it can be useful to write the elements of these groups directly in their fundamental representation. This is a $2N$ dimensional representation where an element $(g,\eta)$ is represented by
\begin{align}\label{eq:fundamental_rep_SUtilde}
    \text{fund}((g,1))=\left(\begin{array}{cc}
            g & 0 \\
            0 & \Theta(-1) (g)
        \end{array} \right)\,,\quad \text{fund}((g,-1)) = 
        \left(\begin{array}{cc}
            0 & g \\
            \Theta(-1)(g) & 0
        \end{array} \right)
\end{align}

With these definitions, it is easy to compute both the center as well as the centralizer of the identity component of these groups. Fisrt, recall that the center of $U(N)$ and $SU(N)$ is $U(1)$ or $\ZZ_N$ respectively, with elements $e^{i\theta}\mathds{1}$ or $e^{i 2k\pi/N}\mathds{1}$. Second, note that elements of the disconnected component don't commute with generic elements in the connected component. Therefore, to find the center, we need to find the elements $(h,1)$ with $h\in Z(G)$ ($G=U(N)$ or $SU(N)$) that commute with $(g,-1)$ with $g\in G$. From the definition of the semidirect product, we find
\begin{align}
   & (h,1)(g,-1) = (h g , -1) \,,\\
&    (g,-1)(h,1) = (g \Theta(-1)(h), -1)\,.
\end{align}
Since $h\in Z(G)$, this leads to the condition
\begin{align}\label{eq:condition_centers}
    h=\Theta(-1)(h)=\overline{h}\,,
\end{align}
which doesn't depend on whether $\Theta$ corresponds to \eqref{eq:semidirect_theta1} or \eqref{eq:semidirect_theta2}. This condition is only satisfied for $h=\pm\mathds{1}$; however, note that for the case of $SU(2k-1)$ only $+\mathds{1}$ belongs to the group. All in all, the center of these groups is given by
\begin{align}\label{eq:centers_SU_tilde}
  &  Z(\widetilde{U}(N))=\ZZ_2\,,\\
  &  Z(\widetilde{SU}(2n))=\ZZ_2\,,\\
  & Z(\widetilde{SU}(2n+1))=\lbrace \mathds{1} \rbrace\,.
\end{align}

The topological GW operators can be found from the centralizers of the identity components. The computation of these centralizers is very similar to that of the centers above. The only difference is that, since these elements don't need to commute with the disconnected component, we don't need to impose \eqref{eq:condition_centers}. Therefore,
\begin{align}\label{eq:centralizers_SU_tilde}
 & C_{\widetilde{U}(N)}(U(N)) = \left\lbrace \left( e^{i \theta} \mathds{1} ,1\right)\,,\, \theta\in [0,2\pi]  \right\rbrace \,,\\
  & C_{\widetilde{SU}(2n)}(SU(2n)) = \left\lbrace \left( e^{i \frac{k\pi}{n}} \mathds{1} ,1\right)\,,\, k=0,1,\dots,2n-1  \right\rbrace \,,\\
   & C_{\widetilde{SU}(2n+1)}(SU(2n+1)) = \left\lbrace \left( e^{i \frac{2k \pi}{2n+1}} \mathds{1} ,1\right)\,,\, k=0,1,\dots,2n  \right\rbrace \,,
\end{align}

\begin{table}[h!]
    \centering
    \begin{tabular}{c|c|c|c}
        Gauge group & Topological GW operators & Quantum dimension & 1-form symmetry \\\hline
        \multirow{3}{*}{$\widetilde{U}(N)$} & $T^{\widetilde{U}}_0=$ Id & 1  &  \multirow{3}{*}{\eqref{eq:fusion_rule_U_tilde}}  \\
                              & $T^{\widetilde{U}}_\theta\,,$ $\theta\in (0,\pi)$ & 2 &  \\
                              & $T^{\widetilde{U}}_\pi$   &  1 & \\\hline
        \multirow{2}{*}{$\widetilde{SU}(2n+1)$ }  &   $T^{\widetilde{SU}}_0=$ Id   &  1  & \multirow{2}{*}{\eqref{eq:fusion_rule_SU_tilde}} \\
           &   $T^{\widetilde{SU}}_k\,,$ $k=1,\dots,n$  &  2  &  \\\hline 
       \multirow{3}{*}{$\widetilde{SU}(2n)$} & $T^{\widetilde{SU}}_0=$ Id & 1  &  \multirow{3}{*}{\eqref{eq:fusion_rule_SU_tilde}}  \\
                              & $T^{\widetilde{SU}}_k\,,$ $k=1,\dots, n-1$ & 2 &  \\
                              & $T^{\widetilde{SU}}_n = T^{\widetilde{SU}}_\pi$   &  1 &
    \end{tabular}
    \caption{Summary of topological Gukov-Witten operators for theories with $\widetilde{U}(N)$ and $\widetilde{SU}(N)$ gauge group.}
    \label{tab:GW_operators_SUtilde_summary}
\end{table}

Finally, in order to find the topological GW operators, what we need to do is compute the conjugacy classes of the elements above in $\widetilde{U}(N)$ and $\widetilde{SU}(N)$. Working directly in the fundamental representation,
\begin{align}
         \left(\begin{array}{cc}
            0 & \mathds{1} \\
            \mathds{1} & 0
        \end{array} \right)
        \left(\begin{array}{cc}
            g & 0 \\
            0 & \Theta(-1)(g)
        \end{array} \right)
        \left(\begin{array}{cc}
            0 & \mathds{1} \\
            \mathds{1} & 0
        \end{array} \right)
        =
        \left(\begin{array}{cc}
            \Theta(-1)(g) & 0 \\
            0 & g
        \end{array} \right)\,.
\end{align}

From the definition of $\Theta(-1)$ we see that conjugating with an element of the disconnected component leads to mapping the phase $\theta\mapsto -\theta$, or $k\mapsto -k$. All in all, for the different gauge groups, we have topological GW operators labelled by
\begin{align}
    \widetilde{U}(N): &\quad T^{\widetilde{U}}_g=T^{\widetilde{U}}_{g(\theta)}\,,\quad g(\theta) = e^{i\theta} \left(\begin{array}{cc}
            \mathds{1} & 0 \\
            0 & \mathds{1}
        \end{array} \right)\,,\, \theta \in [0,\pi]  \,,  \\
    \widetilde{SU}(2n): & \quad T^{\widetilde{SU}}_g = T^{\widetilde{SU}}_{g(k)}\,,\quad g(k) =     e^{i\frac{k\pi}{n}}\left(\begin{array}{cc}
            \mathds{1} & 0 \\
            0 & \mathds{1}
        \end{array} \right)\,,\, k=0,1,\dots,n \,,\\
            \widetilde{SU}(2n+1): & \quad T^{\widetilde{SU}}_g = T^{\widetilde{SU}}_{g(k)}\,,\quad g(k) = e^{i\frac{2k\pi}{2n+1}}\left(\begin{array}{cc}
            \mathds{1} & 0 \\
            0 & \mathds{1}
        \end{array} \right)\,,\, k=0,1,\dots,n\,.
\end{align}

Note that the conjugacy classes corresponding to $\theta \in (0,\pi)$ have two elements belonging to them, and likewise for $k=1,\dots,\lfloor N/2 \rfloor-1$. Therefore the corresponding GW operators have quantum dimension two: this means that the symmetry is non-invertible, as there are no operators that we can fuse with e.g. $T^{\widetilde{SU}}_1$ and produce the identity operator, according to \eqref{eq:quantum_dimension}. This can also be checked by directly computing the fusion rule, which we proceed to illustrate in the example of $\widetilde{SU}(N)$. Note that the GW operator $T^{\widetilde{SU}}_k$ can be written as a sum of two GW operators of $SU$ gauge theory,
\begin{align}
\label{GWtildeSU}
    T^{\widetilde{SU}}_k (M^{d-2})= T^{SU}_k(M^{d-2})+ T^{SU}_{-k}(M^{d-2})\,.
\end{align}
If we fuse two operators with $k\neq k'$, we find
\begin{align}
    T^{\widetilde{SU}}_k \cdot T^{\widetilde{SU}}_{k'} &= \left(T^{SU}_k+ T^{SU}_{-k}\right) \cdot \left(T^{SU}_{k'}+ T^{SU}_{-k'}\right)\\
    & = T^{SU}_k\cdot T^{SU}_{k'} + T^{SU}_k\cdot T^{SU}_{-k'} + T^{SU}_{-k}\cdot T^{SU}_{k'} + T^{SU}_{-k}\cdot T^{SU}_{-k'}\nonumber\,.
\end{align}
The 1-form symmetry of $SU(N)$ gauge theory is invertible, i.e. the fusion of the $T^{SU}_k$ GW operators obeys \eqref{eq:invertible_hfs}. Thus,
\begin{align}\label{eq:fusion_rule_SU_tilde}
    T^{\widetilde{SU}}_k \cdot T^{\widetilde{SU}}_{k'} &= T^{SU}_{k+k'} + T^{SU}_{k-k'} + T^{SU}_{k'-k} + T^{SU}_{-(k+k')} \nonumber \\
    &=T^{\widetilde{SU}}_{k+k'} + T^{\widetilde{SU}}_{|k-k'|}\,,
\end{align}
The fusion rule in the case when $k=k'$ is much more subtle and we cannot compute it explicitly. However, since the behaviour of the 1-form symmetry of $\widetilde{SU}$ and $\widetilde{U}$ groups seems in many aspects a natural generalization of the $O(2)$ case described above, we expect that an analogous argument to that between equations \eqref{eq:O2_weird_fusion_1}--\eqref{eq:O2_weird_fusion_2} should also apply, giving rise to fusion rules similar to \eqref{eq:fusion_algebra_O2}.  

For the gauge group $\widetilde{U}$, the computation is completely analogous, and the only difference is that instead of the discrete parameters $k,k'$ we have continious parameters $\theta, \theta'$,
\begin{align}\label{eq:fusion_rule_U_tilde}
    T^{\widetilde{U}}_{\theta}\cdot T^{\widetilde{U}}_{\theta'} = T^{\widetilde{U}}_{\theta+\theta'} + T^{\widetilde{U}}_{|\theta-\theta'|}\,.
\end{align}
 
We can gain further intuition on the appearance of non-invertible symmetries if we consider the theory before gauging the outer automorphism (i.e. just $U(N)$ or $SU(N)$ gauge theory), when the topological Gukov-Witten operators have quantum dimension one. However, the (now global) zero-form symmetry is not independent of the center one-form symmetry; a feature that can be seen from the fact that the fundamental Wilson line is acted upon by charge conjugation $\mathcal{C}$ as 
\begin{align}
\label{ConWL}
    \mathcal{C}:\,W_{\rm\bf fund} \mapsto W_{\overline{\rm\bf fund}}\,,
\end{align}
This implies that the total global symmetry of the theory is not a direct product of both, but rather a higher categorical object known as a \emph{2-group symmetry} (see \textit{e.g.} \cite{Kapustin:2013uxa,Sharpe:2015mja,Delcamp:2018wlb,Cordova:2018cvg,Benini:2018reh,Delcamp:2019fdp}). 

This observation provides further support to the identification of the symmetry operators for the $\widetilde{SU}(N)/\widetilde{U}(N)$ theories above (along lines similar to \cite{Barkeshli:2014cna}). The GW operators of the $U(N)/SU(N)$ theory are acted in a similar way to \eqref{ConWL} by $\mathcal{C}$. It is then clear that the $\mathcal{C}$-invariant combinations are precisely \eqref{GWtildeSU} (and the converse for $U(N)$), which are then the leftover GW after gauging $\mathcal{C}$. Note that the appearance of non-invertible symmetries is due to the ``folded structure" in \eqref{GWtildeSU}, which from this point of view is inherited from the fact that the 1-form generators of the $SU(N)/U(N)$ theory are acted by the 0-form symmetry $\mathcal{C}$.\footnote{This possibility seems to be excluded in \cite{Casini:2020rgj,Casini:2021zgr} (see footnote 11 in \cite{Casini:2020rgj}).}

\section{Dual $(d-2)$-form symmetry}\label{section:d-2 form}

In this section, the aim is twofold: on the one hand, we study the $(d-2)$-form symmetry generated by topological Wilson lines; and on the other hand, we look for brane constructions of $\widetilde{SU}(N)$ theories. It seems that the key to achieving the second is to include defects so that the $(d-2)$-form symmetry is broken.

\subsection{Wilson lines and Alice strings}

In the previous section we looked for topological GW operators that generate the electric 1-form symmetry. The charged objects were Wilson lines, and the symmetry can be broken if we include particles that make the Wilson lines endable. In this section, we look for topological Wilson lines: the charged objects are precisely the Gukov-Witten operators. As we have discussed above, the Wilson line along a contractible path always belongs to the connected component of the group, and this makes it so that the topological ones, which generate the $(d-2)$-form symmetry, are given by the representations of the gauge group that map the whole identity component $G^0$ to 1 \eqref{eq:dual_symmetry_general}.

In the case of $\widetilde{SU}(N)$ and $\widetilde{U}(N)$, the group of connected components is $\ZZ_2$. This group has two representations, the trivial and the fundamental. At the level of the group, they correspond to the trivial representation and the sign representation respectively, where the sign representation is defined as
\begin{align}\label{eq:sign_rep_SUtilde}
    \text{sign}((g,1))=1\,,\quad \text{sign}((g,-1)) = -1\,,
\end{align}
for $g$ in $\widetilde{SU}(N)$ or $\widetilde{U}(N)$. Therefore, the topological Wilson lines are $W^{\widetilde{G}}_{\mathbf{1}}$ and $W^{\widetilde{G}}_{\text{sign}}$. Similarly to the case of the $O(N)$ groups \eqref{eq:fusion_WL_O_groups}, the topological Wilson lines fuse according to the $\ZZ_2$ product, and therefore we have a $\ZZ_2$ invertible $(d-2)$-form symmetry. 

In a similar fashion to the 1-form symmetry, this $(d-2)$-form symmetry can be broken by including defects on which the charged GW operators can end. These were dubbed \emph{twist vortices} in \cite{Heidenreich:2021xpr}, and in the context of the charge conjugation symmetry have been usually called \emph{Alice strings} \cite{Schwarz:1982ec}. These defects always have a transverse $\mathbb{R}^2$ at each point, and are defined by the fact that when going around them, operators undergo a monodromy corresponding to the outer automorphism of the gauge group.

An important feature of Alice strings is that their presence reduces the globally well defined gauge group \cite{Alford:1992yx}. This is because the outer automorphism doesn't commute with all gauge transformations, and states that should be gauge equivalent can pick up different Aharonov-Bohm phases from the action of the monodromy. This is a contradiction: what happens is that, even if in a region that doesn't include the string the gauge group is apparently $G$, the presence of the string reduces it to the centraliser of the outer automorphism of $G$, i.e. precisely the subgroup that does commute with the monodromy.

Let's be more explicit in the case at hand of $\widetilde{SU}(N)_{I,II}$. When going around the Alice string, fields are acted upon by the element $(1,-1)$ which corresponds to the outer automorphism of $SU(N)$. Note that this action does depend on the choice of semidirect product $\Theta$ \eqref{eq:semidirect_theta1} or \eqref{eq:semidirect_theta2}. The globally well defined gauge group is the centraliser $C_{\widetilde{SU}(N)_{I,II}}((1,-1))$, which can be easily found by computing
\begin{align}
   & (g,\eta) (1,-1) = (g \Theta(\eta)(1),-\eta) = (g,-\eta)\,, \\
   & (1,-1) (g,\eta) = (\Theta(-1)(g),-\eta)\,.
\end{align}
We see that $g$ needs to satisfy
\begin{align}
    g = \Theta(-1) (g)\,.
\end{align}
If $\Theta=\Theta^I$ \eqref{eq:semidirect_theta1}, this implies that $g\in SO(N)$. On the other hand, if $\Theta=\Theta^{II}$ \eqref{eq:semidirect_theta2}, which can only happen if $N$ is even, we have $g\in Sp(N/2)$.

In summary, if we add an Alice string to break the $(d-2)$-form symmetry of $\widetilde{SU}(N)$ theory, the gauge group becomes $SO(N)\times \ZZ_2$ or $Sp(N/2)\times \ZZ_2$. It is interesting to look back to the electric 1-form symmetry of these theories. From Table \ref{tab:GW_operators_SpSO_summary}, we see that the topological Gukov-Witten operators, after the reduction of the well defined gauge group, correspond to $+\mathds{1}$ or $-\mathds{1}$ (the latter only when it belongs to the group). Comparing with the results of Table \ref{tab:GW_operators_SUtilde_summary}, these are precisely the GW operators with quantum dimension 1 in $\widetilde{SU}(N)$ theory. Therefore, it seems that breaking the $(d-2)$-form symmetry ultimately results in the disappearance of the non-invertible 1-from symmetries. We believe that this is a phenomenon deserving of further exploration.

\subsection{Brane constructions}

Brane setups engineering Alice strings and $\widetilde{U}(N)$ gauge groups can be achieved by intersecting branes with orientifolds. Following \cite{Gukov:2008sn}, consider

\begin{align}
        \begin{tabular}{c|cccccccccc}
           & 0 & 1 & 2 & 3 & 4 & 5 & 6 & 7 & 8 & 9  \\\hline
       $N$ D3 & $\times$ & $\times$ & $\times$ & $\times$ &  &   &   &   &   &     \\
        O3 & $\times$ & $\times$ &  &  & $\times$  & $\times$  &   &   &   &     \\
        $k$ D3' & $\times$ & $\times$ &  &  & $\times$ & $\times$  &   &   &   &     
    \end{tabular}
    \label{eq:brane_construction_GW}
\end{align}

The argument in \cite{Gukov:2008sn} suggests that the theory on the D3-branes is $\widetilde{U}(N)$. However, from the point of view of the D3-branes, the O3 $+ k$ D3' act as a codimension 2 defect with a monodromy associated to the element $(1,-1) \in \widetilde{U}(N)$, that is, an Alice string. As a result, the globally well defined gauge group is $O(N)$ (in the $\widetilde{U}(N)_I$ case) or $Sp(N/2)$ (in the $\widetilde{U}(N)_{II}$ case), and the full $\widetilde{U}(N)_{I,II}$ is only manifest on top of the defect. The type of semidirect product extension $\Theta^I$ or $\Theta^{II}$ depends on the choice of orientifold plane. For O3$^+,\,\widetilde{\rm O3}^+$ we find $\widetilde{U}(N)_I$, leading to a globally well-defined $O(N)$ in the presence of the twist vortex, and for an O3$^-$ we find $\widetilde{U}(N)_{II}$ leading to a globally well-defined $Sp(N/2)$ in the presence of the twist vortex.\footnote{It would be interesting to clarify the distiction bewteen O3$^-$ and $\widetilde{\rm O3}^-$.}

This arrangement of branes can be straightforwardly generalised to $N$ D$p$-branes along the $x^{0,\dots,p}$ directions and O$p$ $+k$ D$p$' along the $x^{0,\dots,p-2,p+1,p+2}$. In this way, we can engineer $\widetilde{U}(N)$ theories in a different number of dimensions.

Besides \eqref{eq:brane_construction_GW}, there are two more setups where we can engineer a $\widetilde{U}(N)$ theory in a similar fashion. These are the type IIB configuration

\begin{align}
        \begin{tabular}{c|cccccccccc}
           & 0 & 1 & 2 & 3 & 4 & 5 & 6 & 7 & 8 & 9  \\\hline
 $N$  D3 & $\times$ & $\times$ & $\times$ & $\times$ &  &  &   &   &   &  \\
     $k$  D7' & $\times$ & $\times$ &  &  & $\times$ & $\times$  & $\times$  &  $\times$ &  $\times$ &  $\times$   \\
        O7 & $\times$ & $\times$ & & & $\times$ & $\times$  & $\times$  &  $\times$ & $\times$  & $\times$   
    \end{tabular}
    \label{eq:brane_construction_IIB}
\end{align}
The IIB configuration with an $O7^-$ and 4 extra flavor $D7$ branes was studied in detail in \cite{Harvey:2007ab, Harvey:2008zz}, where it was argued that the 2d intersection acts as an Alice string for the $\widetilde{U}(N)_I$ theory on the D3, out of which only a $O(N)$ is globally well-defined.

In addition, we also have the T-dual in type IIA,

\begin{align}
        \begin{tabular}{c|cccccccccc}
           & 0 & 1 & 2 & 3 & 4 & 5 & 6 & 7 & 8 & 9  \\\hline
 $N$  D2 & $\times$ &  & $\times$ & $\times$ &  &  &   &   &   &  \\
     $k$  D6' & $\times$ &  &  &  & $\times$ & $\times$  & $\times$  &  $\times$ &  $\times$ &  $\times$   \\
        O6 & $\times$ &  & & & $\times$ & $\times$  & $\times$  &  $\times$ & $\times$  & $\times$   
    \end{tabular}
    \label{eq:brane_construction_IIA}
\end{align}

In both these cases, the theory on the D3 (or D2) is a $\widetilde{U}(N)$ gauge theory, and the orientifold intersection appears as an Alice string which reduces the well defined gauge group to an orthogonal or symplectic subgroup. However, as opposed to \eqref{eq:brane_construction_GW}, the identification of the type of orientifold with the semidirect product extension is reversed: an O7$^+$ will give rise to $\widetilde{U}(N)_{II}$ on the D3, and an O7$^-$ to $\widetilde{U}(N)_I$.

An important remark is that all the brane constructions that we have discussed have one thing in common: when seeking to engineer a disconnected gauge group, the Alice strings appears automatically, and it seems impossible to find the former without the latter. As a consequence, as we have discussed in the previous section, the presence of the Alice string also serves the purpose of breaking the $(d-2)$-form global symmetry of these theories. This strongly resonates with the conjectured absence of global symmetries in quantum gravity.

\section{Conclusions and outlook}

In this note we have studied the electric 1-form and $(d-2)$-form symmetries of gauge theories based on the gauge groups which include charge conjugation as part of the gauge symmetry introduced in \cite{Bourget:2018ond,Arias-Tamargo:2019jyh}. As for the electric 1-form symmetry, concentrating on pure gauge theories, we have found that these QFT's provide very simple and explicit examples of non-invertible symmetries using the technology developed in \cite{Heidenreich:2021xpr}, supporting the claim in \cite{Roumpedakis:2022aik} that indeed non-invertible symmetries are ubiquitous also in dimensions higher than three. In this case, the emergence of non-invertible symmetries can be heuristically understood considering a copy of the same theory but with gauge group $SU(N)$ (an analogous discussion holds for the $U(N)$ case). In that case, the symmetry operators associated to the electric 1-form symmetry are permuted by the 0-form charge conjugation symmetry (forming actually a 2-group). The operators which carry over to the version of the theory with gauged charge conjugation are the combinations which are $\mathcal{C}$-invariant, and this ``folds" the GW operators as in \eqref{GWtildeSU} leading to non-invertible symmetries in very much the same way as in the $O(2)$ case discussed in \cite{Heidenreich:2021xpr}. The non-invertible character of the symmetries can be read-off from the fact that their quantum dimension is 2 (instead of 1). This manifests itself also in the fusion rules, which we expect to mimic the $O(2)$ case. Even though we have provided arguments in support of the fusion rules in section \ref{section:electric 1-form}, it would be very interesting to further study this aspect to put them on firmer grounds.

More generally, the existence of non-invertible symmetries has been shown to be closely related to mixed anomalies (see e.g. \cite{Tachikawa:2017gyf,Kaidi:2021xfk}). The fact that, when considering a theory which includes charge conjugation as part of the gauge group, one immediately finds a non-invertible 1-form symmetry, may signal that such a mixed anomaly between the gauge group and its outer automorphism should be present (see \cite{Komargodski:2017dmc} for a related discussion). It would be very interesting to investigate this point further, complementing the studies of \cite{Henning:2021ctv}.

Since the gauge groups we are considering are disconnected ($\pi_0(G)=\mathbb{Z}_2$), QFT's based on them automatically exhibit a $(d-2)$-form symmetry \cite{Heidenreich:2021xpr}. The objects charged under this symmetry are twist vortices (Alice strings in the $d=4$ case). As it is well-known, in the presence of twist vortices only a subgroup of the full gauge group is well defined. In the case at hand it is  $SO(N)$ for $\widetilde{SU}(N)_I$ and $Sp(N/2)$ for $\widetilde{SU}(N)_{II}$ --recall that this later version is only available for even $N$. Elaborating on \cite{Gukov:2008sn} and \cite{Harvey:2007ab, Harvey:2008zz} we have suggested String Theory embeddings for these theories. Amusingly, they automatically come with twist vortices, thus breaking the $(d-2)$-symmetry. This is very much consistent with the Swampland criteria that any global symmetry should be broken. These String Theory constructions involve intersecting orientifolds. Roughly speaking, the type of orientifold ($Op^{\pm}$ and their tilde versions) match the possible theories. However, it would be very interesting to study these constructions in more detail (in particular including the relation between the two proposed constructions). Moreover, the String Theory construction may be used to study the duality properties of these theories. As a consequence, this would allow to study magnetic $(d-3)$-form symmetries as well as possible 't Hooft anomalies. Note that some of these aspects may depend on the dimension $d$. We leave these very interesting aspects for future studies.

\section*{Acknowledgements}

We would like to thank F. Benini, O. Bergman and specially M. Montero for very useful conversations. GAT is supported by the Spanish government schol- arship MCIU-19-FPU18/02221. This work is partly supported by Spanish national grant MINECO-16-FPA2015- 63667-P as well as the Principado de Asturias grant SV-PA-21-AYUD/2021/52177.  
\appendix

\section{Gukov-Witten operators and principal extensions}\label{sec:appendix_Wendt}

In the main text, we have studied the 1-form symmetries of pure gauge theories with disconnected gauge groups, using the known result that the topological Gukov-Witten operators should correspond to the conjugacy classes of elements in the centralizer of the identity component of the group. In this appendix, we find the same topological GW operators by direct computation of their linking with Wilson lines in the adjoint. This computation has two steps: first we identify all possible GW operators, and then we use \eqref{eq:linking_GW_wilson_lines} to find which of them link trivially with the Wilson line; these will be the topological ones.

Gukov-Witten operators where introduced in \cite{Gukov:2006jk,Gukov:2008sn} as codimension two operators that preserve a certain amount of supersymmetry. This was done by finding solutions to Hitchin's equations for the gauge fields, with a singularity at the locus of the operator and prescribed boundary conditions. Said boundary conditions are specified by the monodromy when going around the singular locus, which is an element of the gauge group and whose conjugacy class is a gauge invariant that labels the different Gukov-Witten operators.

If the gauge group is connected, finding all posible GW operators is easy. A basic theorem of Lie theory tells us that in this case, any element of the group is conjugate to an element of the maximal torus, i.e. if we call $G$ the group and $T$ its maximal torus, the map
\begin{align}
    \mathcal{C}:&\, G\times T \to \quad G\\
    & (g,t)\quad\mapsto\quad g^{-1} t g\nonumber
\end{align}
is surjective. For our purposes, this implies that the possible GW operators (topological or not) are labelled by elements of $T$. 

This statement is no longer true if the group is not connected. Still, if we restrict ourselves to the case of principal extensions (namely the group is a semidirect product of its connected component times its group of outer automorphisms), we have lemma 2.1 of \cite{wendt2001weyl}, which is enough for our purposes. The statement in the case where the outer automorphism group is isomorphic to $\ZZ_2$ is that, if we call $G^0$ the identity component of the group and $\Theta$ the map such that $G=G^0\rtimes_\Theta\ZZ_2$, then the map
\begin{align}
    \varphi:&\, G\times T^\Theta \to \quad G^{\text{disc}}\\
    & (g,t)\quad\mapsto\quad g^{-1} \Theta(t) g\nonumber
\end{align}
is surjective onto the disconnected component of the group. Here $G^{\text{disc}}=\Theta \cdot G^0$ denotes said disconnected component, and $T^\Theta$ is the subgroup of the maximal torus of $G^0$ which is left invariant by the action of $\Theta$. Therefore, GW operators specified by a monodromy transformation in the disconnected component of the gauge group can be labelled by elements in $T^\Theta$.

Once all the GW operators have been identified, we look for the ones that link trivially with an adjoint Wilson line: these are the topological ones that generate the 1-form symmetry. From the linking coefficient \eqref{eq:linking_GW_wilson_lines} it follows that we need to solve
\begin{align}
    \chi_{\text{Adj}}(a)=\dim \text{Adj}\,,
\end{align}
where $a\in T$ if we are considering a GW operator in the connected component and $a=\Theta(t),$ $t\in T^\Theta$ if we are considering one in the disconnected component.\\

\textbf{Example: $\widetilde{SU}(3)_I$}

As an example, we can consider the principal extension of $SU(3)$. Its Lie algebra has three positive roots, $\alpha_1$, $\alpha_2$ and $\alpha_1+\alpha_2$, and the outer automorphism exchanges $\alpha_1$ and $\alpha_2$: thus, the invariant subgroup of the torus, $T^\Theta$, precisely corresponds to the root $\alpha_1+\alpha_2$.  In order to write down the characters, it's more convenient to use a modified basis for the fugacities, instead of the usual one, such that the one parameter subgroup corresponding to $\alpha_1+\alpha_2$ is parametrized by $z_1^2$. This can be achieved by selecting fugacities $z_1 z_2^3$ and $z_1/z_2^3$ for the $\alpha_1$ and $\alpha_2$ directions respectively \cite{Arias-Tamargo:2019jyh, Bourget:2018ond}; note that in these terms the action of the outer automorphism is $z_2\mapsto 1/z_2$. With this, the character of the adjoint evaluated in a conjugacy class in the connected and disconnected components gives
\begin{align}
   & \chi_{\text{Adj}}(t) = 2 + z_1^2+\frac{1}{z_1^2} + z_1 z_2^3 + \frac{1}{z_1 z_2^3} + \frac{z_1}{z_2^3} + \frac{z_2^3}{z_1}\,,\quad t\in T \\
   & \chi_{\text{Adj}}(\Theta(t')) = - z_1^2 - \frac{1}{z_1^2}\,,\quad t'\in T^\Theta
\end{align}

The dimension of the adjoint of $\widetilde{SU}(3)_I$ is equal to 8. Since the fugacities are complex numbers of modulus 1, we find that the topological GW operators correspond to elements in the connected component such that $z_1=1$ and $z_2^3=1$, or $z_1=z_2=-1$. The solutions with $z_1=z_2=1$ and $z_1=z_2=-1$ are in fact one and the same, which can be seen from the fact that with this fugacity parametrization, the character of the fundamental is
$\chi_F=z_1 z_2 + 1/z_2^2 + z_2/z_1$. The corresponding Gukov-Witten operator is the identity of the 1-form symmetry. The other two solutions $z_1=1$, $z_2= e^{i\pi/3}$ and $z_1=1$, $z_2=e^{2i\pi/3}$ correspond to different elements of the gauge group, but ones that get identified via conjugation with the generator of the $\ZZ_2$. Therefore, there is one non-trivial GW operator, with quantum dimension two, that generates the 1-form symmetry. This is the same result obtained from the centralizer computation in the main text.

An important remark is that $\chi_{\text{Adj}}(\Theta(t'))=\dim (\text{Adj})$ has no solutions for GW operators corresponding to the disconnected component. This is completely generic and due to the fact that, since $T^\Theta$ has always a smaller dimension than $T$, there will be fewer monomials in the corresponding character than it's needed to have solutions to the equation. Therefore, GW operators labelled by classes in the disconnected component can never be topological.

\bibliographystyle{JHEP}
\bibliography{ref}

\providecommand{\href}[2]{#2}\begingroup\raggedright\begin{thebibliography}{10}

\bibitem{Gaiotto:2014kfa}
D.~Gaiotto, A.~Kapustin, N.~Seiberg and B.~Willett, \emph{{Generalized Global
  Symmetries}}, \href{https://doi.org/10.1007/JHEP02(2015)172}{\emph{JHEP}
  {\bfseries 02} (2015) 172} [\href{https://arxiv.org/abs/1412.5148}{{\ttfamily
  1412.5148}}].

\bibitem{Verlinde:1988sn}
E.~P. Verlinde, \emph{{Fusion Rules and Modular Transformations in 2D Conformal
  Field Theory}},
  \href{https://doi.org/10.1016/0550-3213(88)90603-7}{\emph{Nucl. Phys. B}
  {\bfseries 300} (1988) 360}.

\bibitem{Petkova:2000ip}
V.~B. Petkova and J.~B. Zuber, \emph{{Generalized twisted partition
  functions}}, \href{https://doi.org/10.1016/S0370-2693(01)00276-3}{\emph{Phys.
  Lett. B} {\bfseries 504} (2001) 157}
  [\href{https://arxiv.org/abs/hep-th/0011021}{{\ttfamily hep-th/0011021}}].

\bibitem{Fuchs:2002cm}
J.~Fuchs, I.~Runkel and C.~Schweigert, \emph{{TFT construction of RCFT
  correlators 1. Partition functions}},
  \href{https://doi.org/10.1016/S0550-3213(02)00744-7}{\emph{Nucl. Phys. B}
  {\bfseries 646} (2002) 353}
  [\href{https://arxiv.org/abs/hep-th/0204148}{{\ttfamily hep-th/0204148}}].

\bibitem{Bachas:2004sy}
C.~Bachas and M.~Gaberdiel, \emph{{Loop operators and the Kondo problem}},
  \href{https://doi.org/10.1088/1126-6708/2004/11/065}{\emph{JHEP} {\bfseries
  11} (2004) 065} [\href{https://arxiv.org/abs/hep-th/0411067}{{\ttfamily
  hep-th/0411067}}].

\bibitem{Fuchs:2007tx}
J.~Fuchs, M.~R. Gaberdiel, I.~Runkel and C.~Schweigert, \emph{{Topological
  defects for the free boson CFT}},
  \href{https://doi.org/10.1088/1751-8113/40/37/016}{\emph{J. Phys. A}
  {\bfseries 40} (2007) 11403}
  [\href{https://arxiv.org/abs/0705.3129}{{\ttfamily 0705.3129}}].

\bibitem{Bachas:2009mc}
C.~Bachas and S.~Monnier, \emph{{Defect loops in gauged Wess-Zumino-Witten
  models}}, \href{https://doi.org/10.1007/JHEP02(2010)003}{\emph{JHEP}
  {\bfseries 02} (2010) 003} [\href{https://arxiv.org/abs/0911.1562}{{\ttfamily
  0911.1562}}].

\bibitem{Casini:2020rgj}
H.~Casini, M.~Huerta, J.~M. Magan and D.~Pontello, \emph{{Entropic order
  parameters for the phases of QFT}},
  \href{https://doi.org/10.1007/JHEP04(2021)277}{\emph{JHEP} {\bfseries 04}
  (2021) 277} [\href{https://arxiv.org/abs/2008.11748}{{\ttfamily
  2008.11748}}].

\bibitem{Casini:2021zgr}
H.~Casini and J.~M. Magan, \emph{{On completeness and generalized symmetries in
  quantum field theory}},
  \href{https://doi.org/10.1142/S0217732321300251}{\emph{Mod. Phys. Lett. A}
  {\bfseries 36} (2021) 2130025}
  [\href{https://arxiv.org/abs/2110.11358}{{\ttfamily 2110.11358}}].

\bibitem{Heidenreich:2021xpr}
B.~Heidenreich, J.~McNamara, M.~Montero, M.~Reece, T.~Rudelius and
  I.~Valenzuela, \emph{{Non-Invertible Global Symmetries and Completeness of
  the Spectrum}}, \href{https://doi.org/10.1007/JHEP09(2021)203}{\emph{JHEP}
  {\bfseries 09} (2021) 203}
  [\href{https://arxiv.org/abs/2104.07036}{{\ttfamily 2104.07036}}].

\bibitem{Choi:2021kmx}
Y.~Choi, C.~Cordova, P.-S. Hsin, H.~T. Lam and S.-H. Shao,
  \emph{{Non-Invertible Duality Defects in 3+1 Dimensions}},
  \href{https://arxiv.org/abs/2111.01139}{{\ttfamily 2111.01139}}.

\bibitem{Kaidi:2021xfk}
J.~Kaidi, K.~Ohmori and Y.~Zheng, \emph{{Kramers-Wannier-like Duality Defects
  in (3+1)D Gauge Theories}},
  \href{https://doi.org/10.1103/PhysRevLett.128.111601}{\emph{Phys. Rev. Lett.}
  {\bfseries 128} (2022) 111601}
  [\href{https://arxiv.org/abs/2111.01141}{{\ttfamily 2111.01141}}].

\bibitem{Roumpedakis:2022aik}
K.~Roumpedakis, S.~Seifnashri and S.-H. Shao, \emph{{Higher Gauging and
  Non-invertible Condensation Defects}},
  \href{https://arxiv.org/abs/2204.02407}{{\ttfamily 2204.02407}}.

\bibitem{Rudelius:2020orz}
T.~Rudelius and S.-H. Shao, \emph{{Topological Operators and Completeness of
  Spectrum in Discrete Gauge Theories}},
  \href{https://doi.org/10.1007/JHEP12(2020)172}{\emph{JHEP} {\bfseries 12}
  (2020) 172} [\href{https://arxiv.org/abs/2006.10052}{{\ttfamily
  2006.10052}}].

\bibitem{Palti:2019pca}
E.~Palti, \emph{{The Swampland: Introduction and Review}},
  \href{https://doi.org/10.1002/prop.201900037}{\emph{Fortsch. Phys.}
  {\bfseries 67} (2019) 1900037}
  [\href{https://arxiv.org/abs/1903.06239}{{\ttfamily 1903.06239}}].

\bibitem{vanBeest:2021lhn}
M.~van Beest, J.~Calder\'on-Infante, D.~Mirfendereski and I.~Valenzuela,
  \emph{{Lectures on the Swampland Program in String Compactifications}},
  \href{https://arxiv.org/abs/2102.01111}{{\ttfamily 2102.01111}}.

\bibitem{Bourget:2018ond}
A.~Bourget, A.~Pini and D.~Rodr\'\i{}guez-G\'omez, \emph{{Gauge theories from
  principally extended disconnected gauge groups}},
  \href{https://doi.org/10.1016/j.nuclphysb.2019.02.004}{\emph{Nucl. Phys. B}
  {\bfseries 940} (2019) 351}
  [\href{https://arxiv.org/abs/1804.01108}{{\ttfamily 1804.01108}}].

\bibitem{Arias-Tamargo:2019jyh}
G.~Arias-Tamargo, A.~Bourget, A.~Pini and D.~Rodr\'\i{}guez-G\'omez,
  \emph{{Discrete gauge theories of charge conjugation}},
  \href{https://doi.org/10.1016/j.nuclphysb.2019.114721}{\emph{Nucl. Phys. B}
  {\bfseries 946} (2019) 114721}
  [\href{https://arxiv.org/abs/1903.06662}{{\ttfamily 1903.06662}}].

\bibitem{Schwarz:1982ec}
A.~S. Schwarz, \emph{{FIELD THEORIES WITH NO LOCAL CONSERVATION OF THE ELECTRIC
  CHARGE}}, \href{https://doi.org/10.1016/0550-3213(82)90190-0}{\emph{Nucl.
  Phys. B} {\bfseries 208} (1982) 141}.

\bibitem{Alford:1989ch}
M.~G. Alford, J.~March-Russell and F.~Wilczek, \emph{{Discrete Quantum Hair on
  Black Holes and the Nonabelian {Aharonov-Bohm} Effect}},
  \href{https://doi.org/10.1016/0550-3213(90)90512-C}{\emph{Nucl. Phys. B}
  {\bfseries 337} (1990) 695}.

\bibitem{Bhardwaj:2022yxj}
L.~Bhardwaj, L.~Bottini, S.~Schafer-Nameki and A.~Tiwari, \emph{{Non-Invertible
  Higher-Categorical Symmetries}},
  \href{https://arxiv.org/abs/2204.06564}{{\ttfamily 2204.06564}}.

\bibitem{Gukov:2006jk}
S.~Gukov and E.~Witten, \emph{{Gauge Theory, Ramification, And The Geometric
  Langlands Program}},  \href{https://arxiv.org/abs/hep-th/0612073}{{\ttfamily
  hep-th/0612073}}.

\bibitem{Gukov:2008sn}
S.~Gukov and E.~Witten, \emph{{Rigid Surface Operators}},
  \href{https://doi.org/10.4310/ATMP.2010.v14.n1.a3}{\emph{Adv. Theor. Math.
  Phys.} {\bfseries 14} (2010) 87}
  [\href{https://arxiv.org/abs/0804.1561}{{\ttfamily 0804.1561}}].

\bibitem{Alford:1992yx}
M.~G. Alford, K.-M. Lee, J.~March-Russell and J.~Preskill, \emph{{Quantum field
  theory of nonAbelian strings and vortices}},
  \href{https://doi.org/10.1016/0550-3213(92)90468-Q}{\emph{Nucl. Phys. B}
  {\bfseries 384} (1992) 251}
  [\href{https://arxiv.org/abs/hep-th/9112038}{{\ttfamily hep-th/9112038}}].

\bibitem{Henning:2021ctv}
B.~Henning, X.~Lu, T.~Melia and H.~Murayama, \emph{{Outer automorphism
  anomalies}}, \href{https://doi.org/10.1007/JHEP02(2022)094}{\emph{JHEP}
  {\bfseries 02} (2022) 094}
  [\href{https://arxiv.org/abs/2111.04728}{{\ttfamily 2111.04728}}].

\bibitem{Kapustin:2013uxa}
A.~Kapustin and R.~Thorngren, \emph{{Higher symmetry and gapped phases of gauge
  theories}},  \href{https://arxiv.org/abs/1309.4721}{{\ttfamily 1309.4721}}.

\bibitem{Sharpe:2015mja}
E.~Sharpe, \emph{{Notes on generalized global symmetries in QFT}},
  \href{https://doi.org/10.1002/prop.201500048}{\emph{Fortsch. Phys.}
  {\bfseries 63} (2015) 659}
  [\href{https://arxiv.org/abs/1508.04770}{{\ttfamily 1508.04770}}].

\bibitem{Delcamp:2018wlb}
C.~Delcamp and A.~Tiwari, \emph{{From gauge to higher gauge models of
  topological phases}},
  \href{https://doi.org/10.1007/JHEP10(2018)049}{\emph{JHEP} {\bfseries 10}
  (2018) 049} [\href{https://arxiv.org/abs/1802.10104}{{\ttfamily
  1802.10104}}].

\bibitem{Cordova:2018cvg}
C.~C\'ordova, T.~T. Dumitrescu and K.~Intriligator, \emph{{Exploring 2-Group
  Global Symmetries}},
  \href{https://doi.org/10.1007/JHEP02(2019)184}{\emph{JHEP} {\bfseries 02}
  (2019) 184} [\href{https://arxiv.org/abs/1802.04790}{{\ttfamily
  1802.04790}}].

\bibitem{Benini:2018reh}
F.~Benini, C.~C\'ordova and P.-S. Hsin, \emph{{On 2-Group Global Symmetries and
  their Anomalies}}, \href{https://doi.org/10.1007/JHEP03(2019)118}{\emph{JHEP}
  {\bfseries 03} (2019) 118}
  [\href{https://arxiv.org/abs/1803.09336}{{\ttfamily 1803.09336}}].

\bibitem{Delcamp:2019fdp}
C.~Delcamp and A.~Tiwari, \emph{{On 2-form gauge models of topological
  phases}}, \href{https://doi.org/10.1007/JHEP05(2019)064}{\emph{JHEP}
  {\bfseries 05} (2019) 064}
  [\href{https://arxiv.org/abs/1901.02249}{{\ttfamily 1901.02249}}].

\bibitem{Barkeshli:2014cna}
M.~Barkeshli, P.~Bonderson, M.~Cheng and Z.~Wang, \emph{{Symmetry
  Fractionalization, Defects, and Gauging of Topological Phases}},
  \href{https://doi.org/10.1103/PhysRevB.100.115147}{\emph{Phys. Rev. B}
  {\bfseries 100} (2019) 115147}
  [\href{https://arxiv.org/abs/1410.4540}{{\ttfamily 1410.4540}}].

\bibitem{Harvey:2007ab}
J.~A. Harvey and A.~B. Royston, \emph{{Localized modes at a D-brane-O-plane
  intersection and heterotic Alice atrings}},
  \href{https://doi.org/10.1088/1126-6708/2008/04/018}{\emph{JHEP} {\bfseries
  04} (2008) 018} [\href{https://arxiv.org/abs/0709.1482}{{\ttfamily
  0709.1482}}].

\bibitem{Harvey:2008zz}
J.~A. Harvey and A.~B. Royston, \emph{{Gauge/Gravity duality with a chiral
  N=(0,8) string defect}},
  \href{https://doi.org/10.1088/1126-6708/2008/08/006}{\emph{JHEP} {\bfseries
  08} (2008) 006} [\href{https://arxiv.org/abs/0804.2854}{{\ttfamily
  0804.2854}}].

\bibitem{Tachikawa:2017gyf}
Y.~Tachikawa, \emph{{On gauging finite subgroups}},
  \href{https://doi.org/10.21468/SciPostPhys.8.1.015}{\emph{SciPost Phys.}
  {\bfseries 8} (2020) 015} [\href{https://arxiv.org/abs/1712.09542}{{\ttfamily
  1712.09542}}].

\bibitem{Komargodski:2017dmc}
Z.~Komargodski, A.~Sharon, R.~Thorngren and X.~Zhou, \emph{{Comments on Abelian
  Higgs Models and Persistent Order}},
  \href{https://doi.org/10.21468/SciPostPhys.6.1.003}{\emph{SciPost Phys.}
  {\bfseries 6} (2019) 003} [\href{https://arxiv.org/abs/1705.04786}{{\ttfamily
  1705.04786}}].

\bibitem{wendt2001weyl}
R.~Wendt, \emph{Weyl's character formula for non-connected lie groups and
  orbital theory for twisted affine lie algebras}, {\emph{Journal of Functional
  Analysis} {\bfseries 180} (2001) 31}.

\end{thebibliography}\endgroup

\end{document}